\documentclass[a4paper,oldversion,letter]{aa}
\usepackage{graphics,txfonts}

\usepackage{natbib}
\bibpunct[, ]{(}{)}{;}{a}{}{,}

\newcommand{\teff}{$T_{\rm eff}$}
\newcommand{\lgg}{$\log g$}

\newcommand{\vs}{$v_{\rm e}\sin i$}
\newcommand{\bs}{$\langle B \rangle$}
\newcommand{\kms}{km\,s$^{-1}$}
\newcommand{\ms}{m\,s$^{-1}$}
\newcommand{\hd}{HD\,75445}
\newcommand{\equ}{$\gamma$\,Equ}
\newcommand{\pr}{\ion{Pr}{iii}}
\newcommand{\nd}{\ion{Nd}{iii}}
\newcommand{\ndt}{\ion{Nd}{ii}}

\newcommand{\fifps}[2]{\centering\resizebox{#1}{!}{\includegraphics{#2}}}

\newcommand{\beq}{\begin{equation}}
\newcommand{\eeq}{\end{equation}}

\begin{document}

\title{Discovery of very low amplitude 9-minute multiperiodic \\ pulsations in the magnetic Ap star HD\,75445%
\thanks{Based on observations collected at the European Southern Observatory,
Chile (ESO programs 68.D-0254, 079.D-0118)}}

%\author{O. Kochukhov   \inst{1} \and
%        S. Bagnulo     \inst{2} \and
%        G. Lo Curto    \inst{3}}
\author{O. Kochukhov   \inst{1} \and
        S. Bagnulo     \inst{2} \and
        G. Lo Curto    \inst{3} \and
        T. Ryabchikova \inst{4,5}}

\offprints{O. Kochukhov, \email{oleg@fysast.uu.se}}

\institute{Department of Physics and Astronomy, Uppsala University, SE-751 20, Uppsala, Sweden
      \and Armagh Observatory, College Hill, Armagh BT61 9DG, Northern Ireland, UK
      \and European Southern Observatory, Alonso de Cordova 3107, 
           Vitacura, Santiago, Casilla 19001 Santiago 19, Chile}
\institute{Department of Physics and Astronomy, Uppsala University, SE-751 20, Uppsala, Sweden
      \and Armagh Observatory, College Hill, Armagh BT61 9DG, Northern Ireland, UK
      \and European Southern Observatory, Alonso de Cordova 3107, 
           Vitacura, Santiago, Casilla 19001 Santiago 19, Chile
      \and Institute of Astronomy, Russian Academy of Sciences, Pyatnitskaya 48, 109017 Moscow, Russia
      \and Department of Astronomy, University of Vienna, T\"urkenschanzstra{\ss}e 17, 1180
      Vienna, Austria}

\date{Received 25 November 2008 / Accepted 08 December 2008}

\abstract{
We present our discovery of pulsational radial-velocity variations in the cool Ap star \hd, an object
spectroscopically similar to the bright, rapidly-oscillating Ap (roAp) star \equ. Based on high-resolution
time-series spectroscopy obtained with the HARPS spectrometer at the European Southern Observatory 3.6-m telescope,
we detected oscillations in \ndt\ and \nd\ lines with a period close to 9~min and amplitudes of 20--30~\ms.
Substantial variation in the pulsational amplitude during our 3.8~h observing run reveals the presence of
at least three excited non-radial modes. The detection of extremely low amplitude pulsations in \hd\ indicates that
the roAp excitation mechanism produces variability in the radial velocity amplitude of between a few
tens \ms\ and several \kms. This supports the idea that many, if not all, cool Ap stars occupying the roAp
instability strip may harbour non-radial pulsations, which currently remain undetected due to their small
photometric and radial-velocity amplitudes.
}


\maketitle

\section{Introduction}
\label{intro}

Rapidly-oscillating Ap (roAp) stars are magnetic chemically-peculiar stars pulsating in non-radial,
magnetoacoustic {\it p-}modes of periods close to 10~min. Pulsations in roAp stars are believed to be
driven by the opacity mechanism operating in the hydrogen-ionization zone
\citep{balmforth:2001}. The presence of strong magnetic fields in roAp stars enhances the
driving of the high-overtone oscillations by the suppression of convection, influences pulsation
frequencies, and determines the global geometry of the pulsational perturbations
\citep{dziembowski:1996,saio:2005}. 

Strongly inhomogeneous vertical distributions of chemical elements combined with the rapid
transformation of the outwardly-propagating pulsation waves are responsible for the unique
spectroscopic pulsation signature of roAp stars \citep{ryabchikova:2002}. In particular, the
lines of rare-earth elements (REEs) and the cores of the hydrogen lines pulsate with a factor of 10--100
higher amplitudes than the remaining spectral features
\citep{kochukhov:2001b,mkrtichian:2003,kurtz:2006b}.

The observed roAp instability strip is \textbf{limited} to the \teff\ range 6400--8100~K, although theoretical
stability calculations \citep{cunha:2002} predict pulsations in stars as hot as 9500~K and are unable to
account for the unstable modes observed in stars with \teff\,$\le$\,7400~K. At the same time, the coexistence
of pulsating and apparently constant Ap stars in the same region of the H-R diagram has been a long-standing
puzzle \citep[e.g.,][]{martinez:1994}. It is now understood that the time-resolved photometric techniques
employed to detect and study variability in most of the 37 known roAp stars are insensitive to the
low-amplitude pulsations observed in spectroscopic time-series analyses
\citep{hatzes:2004,elkin:2005}. This leads to the suggestion that all magnetic Ap stars in a certain
temperature range may oscillate, but some have amplitudes below the photometric detection threshold
\citep{kochukhov:2002}. 

We test this hypothesis by completing a high-precision survey of a sample of bright cool magnetic Ap stars
using the  High Accuracy Radial velocity Planet Searcher (HARPS) spectrograph at the European Southern
Observatory (ESO). The first result of our observations -- the discovery of 10.9-minute oscillations in the
Ap star HD\,115226 -- was reported by \citet{kochukhov:2008b}. Here we present the discovery of a new
rapidly oscillating Ap star, \object{\hd}, which pulsates with one of the lowest radial-velocity (RV) amplitudes measured for roAp stars.

\section{Observations and data reduction}
\label{obs}

We used the HARPS spectrograph \citep{mayor:2003} at the ESO 3.6-m telescope at La Silla to monitor \hd\ in
the context of our search for low-amplitude variability in bright cool Ap stars (ESO program 079.D-0118). The star
was observed on the  night of April 15, 2007. The observations started at the barycentric JD 2454205.47129 and
continued for 3.8 hours. We collected 120 consecutive 80~s exposures, separated by a dead time of 31~s. The
resulting time resolution of 111~s enabled us to detect variations \textbf{with} frequencies as high as $\nu=4.5$~mHz
($P=3.7$~min).

The extraction of one-dimensional spectra and barycentric velocity correction of the wavelength scale was
performed with the help of the HARPS pipeline. Our spectra have a nominal resolving power
$R\equiv\lambda/\Delta\lambda=115\,000$, and cover a wavelength range from 3780 to 6910~\AA, with a 30~\AA\ gap close to
5320~\AA. Individual exposures of \hd\ have peak signal-to-noise ratio of 70 per 15~m\AA\ pixel
at $\lambda$~6000~\AA.
In the final reduction step, one-dimensional extracted spectra of \hd\ were post-processed to achieve
consistent continuum normalization following the procedure described in \citet{kochukhov:2007}.

%For the moderate signal-to-noise ratio of individual spectra of \hd\ the dominant source of noise in the radial
%velocity measurements is the photon noise rather than the instrumental precision. For this reason, we did not
%employ the simultaneous ThAr method available at HARPS, avoiding contamination of the stellar signal. Instead, 
%we took a ThAr reference spectrum at the beginning and at the end of the stellar observations. Using these
%calibrations, we estimate that the instrumental drift within the time series was below 0.1~\ms.

We did not employ the simultaneous ThAr method available at HARPS, avoiding contamination of the stellar
signal. Instead, we acquired a ThAr reference spectrum at the beginning and end of the stellar observations.
Using these calibrations, we estimated that the instrumental drift within the time series was below 0.1~\ms. For
the moderate signal-to-noise ratio of individual spectra of HD 75445, the dominant source of noise in the radial-velocity measurements was photon noise ($\ge$\,2~\ms) rather than the instrumental precision, which is similar to the
measured drift.
%: 0.1m/s.

%drift first: -0.57, drift last: -0.53 (first of HD 115226 which followed HD 75445)

\section{Basic properties of HD\,75445}
\label{params}

The southern chemically-peculiar star \hd\ (HIP\,43257, CD $-38^{\rm o}$4907) was classified as a Sr-Eu object by
\citet{bidelman:1973}. Its Str\"omgren photometric indices, $b-y=0.159$, $m_1=0.218$, $c_1=0.729$
\citep{vogt:1979}, H$\beta=2.801$ \citep{Maitzen:2000}, indicate \teff\,=\,7600--7700~K according to the
calibrations by \citet{moon:1985} and \citet{napiwotzki:1993}. Geneva colours yield \teff\,=\,7680~K
\citep{kochukhov:2006}, in good agreement with the Str\"omgren photometry.

\citet{kochukhov:2006} investigated the evolutionary state of \hd\ using Hipparcos parallax and photometric \teff.
They determined $\log{L}=1.17\pm0.06$\,$L_\odot$, $M=1.81\pm0.05$\,$M_\odot$ and a stellar age that is a factor 0.56--0.72 of
the main-sequence lifetime. \citet{ryabchikova:2004a} included \hd\ in their abundance analysis study of a sample of
roAp and non-pulsating Ap stars. Adopting \teff\,=\,7700~K and \lgg\,=\,4.3, they showed that \hd\ has close
to solar Fe abundance, moderate enhancement of Cr and Mn, 1.6~dex overabundance of Co, and a large
overabundance of several REEs. As for many known roAp stars, \hd\ exhibits an ionization anomaly of Pr and Nd,
with doubly ionized lines of these elements providing an 1.3--2.0~dex higher abundance measurement than the lines of first ions.
\citet{ryabchikova:2008} examined the Ca stratification and isotopic composition of \hd. They reported a 2.0~dex 
step-like change of the Ca concentration at $\log\tau_{5000}=-0.9$ and detected the presence of heavy Ca isotopes
($^{46}$Ca and $^{48}$Ca) in the upper atmospheric layers. This study also inferred a spectroscopic \teff\,=\,7650~K
using the H$\alpha$ line.

\begin{figure}[!t]
%\figps{11419_f1.eps}
\fifps{8.5cm}{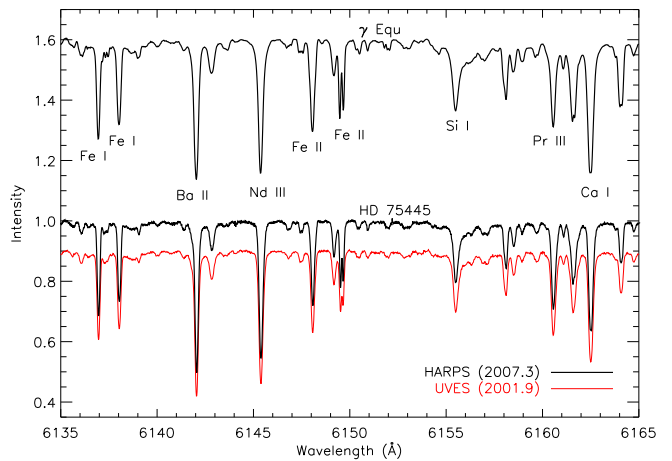}
\caption{Comparison of the 6135--6165~\AA\ region in the spectra of the new roAp star \hd\ and the
well-known bright roAp star \equ. The UVES spectrum of \equ\ is shown on top, with the
identifications of the strongest spectral features. The mean
HARPS spectrum of \hd\ (middle, thick curve) is compared with the UVES observation of this star
(bottom, thin curve) obtained 5.4 years before the HARPS observations. The UVES spectra are shifted
vertically for display purposes.
}
\label{fig1}
\end{figure}

\citet{ryabchikova:2004a} commented on the spectroscopic similarity of \hd\ and the bright roAp star \equ\
(\object{HD\,201601}). This point is illustrated in Fig.~\ref{fig1} with new high-quality, $R=115\,000$
spectra available for both stars (mean HARPS spectrum for \hd\ and mean UVES spectrum derived from the
archival time series data set of \equ). The spectra of these two stars are almost identical, the only
difference being slightly broader line profiles of \equ\ due to the stronger mean surface field strength of
this star. However, there is a discrepancy between the spectra of the two stars in the region of the resonance \ion{Li}{i}
doublet at $\lambda$ 6708~\AA, which is strong in \equ\ but entirely absent in \hd\ \citep{kochukhov:2008a}.

\citet{mathys:1997b} detected Zeeman splitting in the \ion{Fe}{ii} 6149~\AA\ line of \hd\ and measured the mean
field modulus \bs\,=\,$2985\pm42$~G with 9 spectra recorded over the period of 450~d in 1994--1995.
\citet{ryabchikova:2004a} provided three additional measurements of \bs, 2915, 2957, and 2873~G, for the spectra obtained
in 2000--2001. The splitting of \ion{Fe}{ii} 6149~\AA\ in our mean HARPS spectrum (2007.3) and in the UVES
spectrum from 2001 \citep{ryabchikova:2008} is consistent with \bs\,=\,3030~G. In summary, the full set of 14 \bs\
measurements shows no evidence of periodic field strength variation.

\begin{figure}[!t]
%\figps{11419_f2.eps}
\fifps{8.5cm}{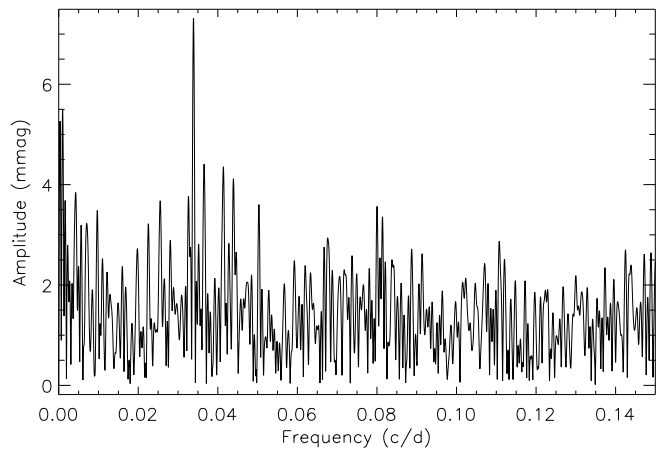}
\caption{Amplitude spectrum for the ASAS photometry of \hd\ obtained after 2003.6.}
\label{fig2}
\end{figure}

We searched for long-period rotational brightness modulation in \hd\ using the Hipparcos epoch
photometry \citep{ESA:1997} and the ASAS database \citep{pojmanski:2002}. No periodic variability \textbf{with} an
amplitude larger than 5~mmag was detectable in the Hipparcos light curve. The ASAS photometry of \hd\ exhibited erratic
brightness changes before 2003, which were probably instrumental in nature. Measurements obtained after
2003.6 did not deviate significantly from the mean value $V=7.14$. The amplitude spectrum computed for the ASAS
observations of \hd\ during 2003.6--2008.8 shows a marginal 7~mmag variability with a 29.5~d period
(Fig.~\ref{fig2}). This is consistent with our estimate of \vs\,$\le$\,2~\kms\ obtained by fitting profiles
of magnetically insensitive \ion{Fe}{i} lines at $\lambda$ 5434, 5576, and 5691~\AA. Comparison of the mean
HARPS spectrum with the UVES observation obtained 5.4 years before our observing run shows no detectable
changes in the line profiles (Fig.~\ref{fig1}), suggesting a very long rotation period.

Spectroscopic similarity of \hd\ to \equ, its prominent REE ionization anomaly, and
its effective temperature of \teff\,$<$\,8000~K imply that
this star is an obvious candidate for the search of oscillations \citep{ryabchikova:2004a}. However,
no photometric pulsation signature exceeding one mmag was detected for this star by Martinez (private
communication). Here, we demonstrate that \hd\ is indeed a roAp star but pulsating with an amplitude well below the
current detection threshold of the ground-based, time-resolved photometry.

\section{Analysis of radial velocity variation}
\label{rv}

We measured radial velocities of lines in the spectrum of \hd\ using the centre-of-gravity technique
\citep{kochukhov:2001b}. Spectral line identification was based on the atomic line data extracted from the
VALD
%\footnote{{http://www.astro.uu.se/~vald/}} 
database \citep{kupka:1999}, which includes the
DREAM
%\footnote{{http://w3.umh.ac.be/~astro/dream.shtml}} 
compilation of the REE line parameters
\citep{biemont:1999}. The list of \nd\ transitions was further extended using the study by
\citet{ryabchikova:2006}.

Previous time-resolved spectroscopic analyses of roAp stars
\citep{kochukhov:2001b,mkrtichian:2003,ryabchikova:2007b} demonstrated that
maximum pulsation amplitudes are always found in singly and doubly ionized REE absorption
features, such as \ndt, \nd, and \pr. A number of strong and medium-strength lines of REE
ions are present in the spectrum of \hd.  However, observational data available to us were of insufficiently high signal-to-noise ratio to detect pulsations in individual lines. We 
reduced the noise in the velocity curves by averaging RV measurements for all lines of \textbf{a given
REE ion}. Among rare-earths, only \ndt\ and \nd\ have sufficient number of lines in the
spectrum of \hd\ to yield precise combined RV measurements. 

\begin{figure}[!th]
%\figps{11419_f3.eps}
\fifps{8.5cm}{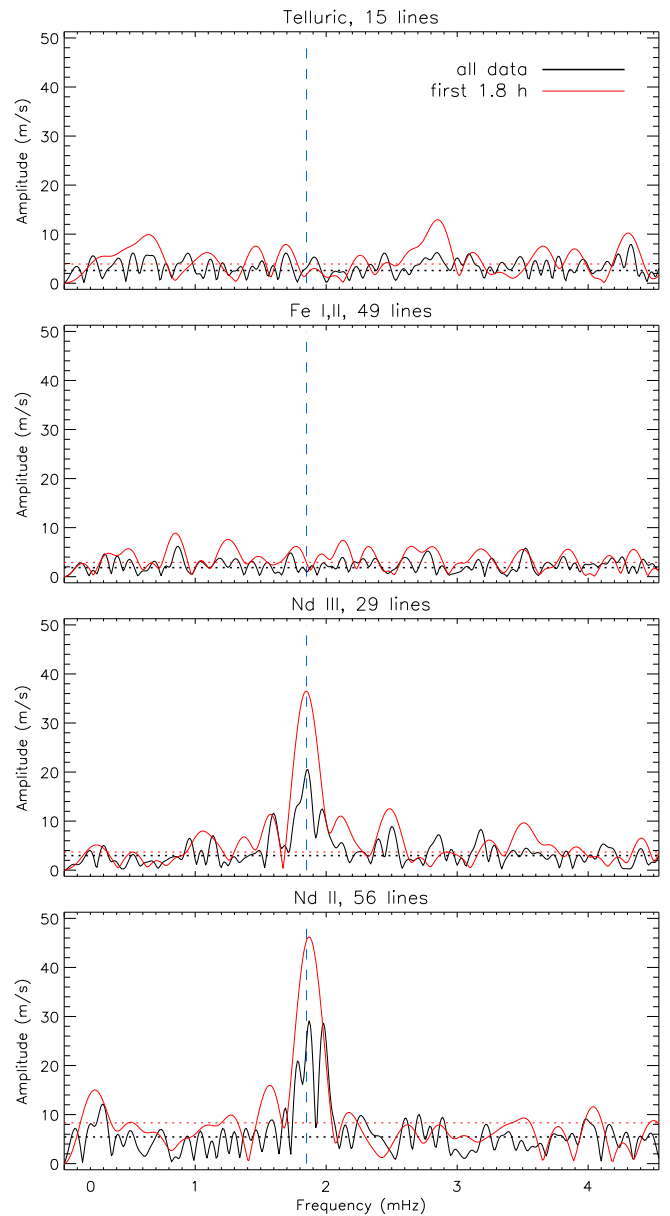}
\caption{From the top panel to the bottom panel: amplitude spectra for the average 
radial-velocity curves of 15 telluric lines, 
49 lines of \ion{Fe}{i} and {\sc ii}, 29 lines of \nd, and 56 lines of \ndt. The vertical
dashed line shows the main pulsation frequency $\nu=1.85$~mHz. In each panel, 
the amplitude spectrum of the entire data set (thick curve) is compared to that of the first 1.8~h of the
spectroscopic monitoring (thin curve). Horizontal dotted lines show corresponding noise levels.}
\label{fig3}
\end{figure}

Using 29 lines of \nd\ and 56 lines of \ndt, we achieved a noise level of 3--5~\ms\ in the
amplitude spectra and revealed conspicuous amplitude peaks in the 1.8--2.0~mHz frequency range, which implied that oscillations of amplitude 20--30~\ms\ were present
(Fig~\ref{fig3}, Table~\ref{tbl1}). These oscillation signatures are highly significant. The 
probability that noise would produce a peak of this observed amplitude at \textit{any} frequency
in the studied range \citep[False Alarm Probability,][]{horne:1986} is $7\times10^{-5}$
for \ndt\ and $4\times10^{-6}$ for \nd. We also applied a bootstrap randomization technique
\citep{kuerster:1997}, which is a more rigorous method of establishing the statistical
significance of a peak in amplitude spectrum. Of the $10^5$ randomly shuffled data sets 
created from the original mean \ndt\ and \nd\ RV curves, none exhibited spurious peaks of the
observed amplitude in the frequency range 0--4.5~mHz. Thus, the probability that noise would
create the signal detected in \ndt\ and \nd\ lines is less than $10^{-5}$.

Applying a similar analysis procedure to 15 telluric lines in the 6275--6315~\AA\ region, we
find no oscillations above 8~\ms\ with a noise level of 3~\ms. Similarly, for the combined RV
curve of 49 \ion{Fe}{i} and \ion{Fe}{ii} lines, which are unexpected to show significant variation
in a low-amplitude roAp star, we found a maximum amplitude of 6~\ms\ and a noise level of 2~\ms.
The stability of telluric lines and the stellar Fe features confirms that variation detected in the Nd
lines of \hd\ is not due to an instrumental artifact.

\begin{table}[!t]
\centering
\caption{Frequency analysis of the average radial velocity curves of telluric lines, \ion{Fe}{i} and {\sc ii},
\nd\ and \ndt. $N$ indicates the number of lines measured. $A_{\rm max}$ gives the highest radial velocity 
amplitude, followed by the estimate of False Alarm Probability of the corresponding signal. The last
two columns give the amplitude of the variation with $\nu=1.85$~mHz ($P=9.01$~min) and the noise estimate.
\label{tbl1}}
\begin{tabular}{lccccc}
\hline
\hline
Ion                  & N  & $A_{\rm max}$ & FAP & $A$ & $\sigma$ \\
                     &    & (\ms)         &     & (\ms)        & (\ms)    \\
\hline
\multicolumn{6}{c}{Full data set}\\
Telluric             & 15 &  7.9 & 0.31    &  $2.5\pm2.5$ & 2.6 \\
\ion{Fe}{i},{\sc ii} & 49 &  6.2 & 0.31    &  $1.5\pm1.8$ & 1.8 \\
\ndt\                & 56 & 29.1 & 6.6E-05 & $24.6\pm4.8$ & 4.9 \\
\nd\                 & 29 & 20.5 & 3.6E-06 & $20.4\pm3.0$ & 3.0 \\
\multicolumn{6}{c}{First 1.8~h}\\
\ndt\                & 56 & 46.2 & 8.7E-06 & $44.2\pm5.5$ & 5.5 \\
\nd\                 & 29 & 36.5 & 1.1E-06 & $36.4\pm3.8$ & 3.7 \\
\hline
\end{tabular}
\end{table}

The complex appearance of the \ndt\ and \nd\ amplitude spectra in Fig.~\ref{fig3} suggests a multiperiodic
pulsation. The presence of several excited modes in \hd\ became apparent when the mean RV data was analysed in
the time domain. We found that in the first 55 observations of \hd, corresponding to the initial 1.8~h of our
time-resolved observations, pulsation variability was clearly evident. The amplitude spectra of this partial
data set, illustrated in Fig.~\ref{fig3}, indicated an almost monoperiodic pulsation. The least-squares fitting
of this part of the oscillation curve yielded amplitudes of 36 and 46~\ms\ as well as pulsation periods of
$8.93\pm0.05$~min and $9.04\pm0.04$~min for the singly and doubly ionized Nd, respectively. There was a small
lag of $0.44\pm0.19$~rad between the RV maxima of the two Nd ions. In a similar way to other roAp stars
\citep{ryabchikova:2007b}, \nd\ in \hd\ showed a later maximum than \ndt.

The prominent sinusoidal variation in the Nd lines was subdued after about 2~h from the beginning of our observations, presumably
due to beating of several excited modes as seen in other multiperiodic roAp stars \citep[e.g.,][]{sachkov:2008}. Tentative
least-squares analysis suggested the presence of at least three significant frequencies: $\nu_1=1.81$~mHz
($P_1=9.20$~min), $\nu_2=1.85$~mHz ($P_2=9.01$~min), and $\nu_3=1.99$~mHz ($P_3=8.37$~min). The length of
our time series does not allow $\nu_1$ and $\nu_2$ to be fully resolved. On the other hand, $\nu_2$ and $\nu_3$
were resolved, and $\nu_3$ was found to have a higher relative amplitude for \ndt.

\section{Discussion}
\label{discus}

We have established the presence of multiperiodic pulsations in the cool magnetic Ap star \hd\
using combined RV measurements of the lines belonging to \ndt\ and \nd. The star exhibits oscillations
with three frequencies, which have different amplitude ratios for the two Nd ions. The phase lag between RV
curves of \ndt\ and \nd\ can be interpreted in the framework of the outwardly propagating pulsational
perturbation, which first reaches the layer where \ndt\ lines form and, after some delay, is seen in the
higher atmospheric layer probed by stronger \nd\ lines \citep{ryabchikova:2007b,mashonkina:2005}. The
difference in the amplitude ratios of the frequency components of the \ndt\ and \nd\ RV curves can be
ascribed to different vertical cross-sections of the three pulsation modes.

\hd\ is the 38th known roAp star. Its discovery is significant because the star's pulsation amplitude is
noticeably lower than for other roAp stars discovered to date using time-resolved spectroscopy. For
example, HD\,218994 \citep{gonzalez:2007} and HD\,115226 \citep{kochukhov:2008b} pulsate with an amplitude 
$\ge$\,500~\ms, while for HD\,116114 \citep{elkin:2005} and HD\,154708 \citep{kurtz:2006b} pulsations with
an amplitude of 50--100~\ms\ were reported. Only for $\beta$\,CrB (HD\,137909) comparable RV amplitudes of
20--30~\ms\ were found in individual lines of singly ionized REEs \citep{kurtz:2007a}. However, $\beta$\,CrB
is in many respects different from other roAp stars. It is an evolved star with a long pulsation period and
a chemical composition deviating from that of a typical roAp star \citep{ryabchikova:2004a}. In contrast, \hd\
appears to have average roAp characteristics and is, in fact, a spectroscopic twin of the well-known roAp
star \equ. Nevertheless, it pulsates with an unusually low RV amplitude. This shows that although the
atmospheric chemical composition, in particular the REE ionization anomaly, is helpful in selecting roAp 
candidates, it has no direct connection with the amplitude of oscillations in the line-forming
region.

\citet{kurtz:2006b} noted the tendency for weaker roAp oscillations to be found in stars with stronger fields.
However, \hd\ has a mean field modulus that is significantly weaker than many roAp stars but also shows an exceptionally low
pulsation amplitude. We therefore conclude that a low-amplitude roAp pulsation can be present in cool Ap stars
of any field strength. Although there are theoretical reasons to believe that the magnetic field alters the
amplitude of the photospheric oscillations in few strong-field stars, a parameter other than the field intensity
defines pulsation amplitude for other roAp stars.

The detection of \textbf{very} low amplitude pulsations in \hd\ suggests that the roAp excitation mechanism produces
oscillations with no apparent lower amplitude threshold. Thus, many cool Ap stars may possess pulsations
with RV amplitudes $\ll$\,100~\ms, which can be currently detected by time-resolved spectroscopy only in
bright sharp-line stars such as \hd.

%\begin{acknowledgements}
%The authors thank Dr. T. Ryabchikova for useful discussions.
%\end{acknowledgements}

%\bibliographystyle{aa}
%\bibliography{astro_papers}

\begin{thebibliography}{38}
\expandafter\ifx\csname natexlab\endcsname\relax\def\natexlab#1{#1}\fi

\bibitem[{{Balmforth} {et~al.}(2001){Balmforth}, {Cunha}, {Dolez}, {Gough}, \&
  {Vauclair}}]{balmforth:2001}
{Balmforth}, N.~J., {Cunha}, M.~S., {Dolez}, N., {Gough}, D.~O., \& {Vauclair},
  S. 2001, \mnras, 323, 362

\bibitem[{{Bidelman} \& {MacConnell}(1973)}]{bidelman:1973}
{Bidelman}, W.~P. \& {MacConnell}, D.~J. 1973, \aj, 78, 687

\bibitem[{{Bi{\'e}mont} {et~al.}(1999){Bi{\'e}mont}, {Palmeri}, \&
  {Quinet}}]{biemont:1999}
{Bi{\'e}mont}, E., {Palmeri}, P., \& {Quinet}, P. 1999, \apss, 269, 635

\bibitem[{{Cunha}(2002)}]{cunha:2002}
{Cunha}, M.~S. 2002, \mnras, 333, 47

\bibitem[{{Dziembowski} \& {Goode}(1996)}]{dziembowski:1996}
{Dziembowski}, W.~A. \& {Goode}, P.~R. 1996, \apj, 458, 338

\bibitem[{{Elkin} {et~al.}(2005){Elkin}, {Riley}, {Cunha}, {Kurtz}, \&
  {Mathys}}]{elkin:2005}
{Elkin}, V.~G., {Riley}, J.~D., {Cunha}, M.~S., {Kurtz}, D.~W., \& {Mathys}, G.
  2005, \mnras, 358, 665

\bibitem[{ESA(1997)}]{ESA:1997}
ESA. 1997, ESA Special Publication, Vol. 1200, {The HIPPARCOS and TYCHO
  catalogues}

\bibitem[{{Gonzalez} {et~al.}(2007){Gonzalez}, {Hubrig}, \&
  {Savanov}}]{gonzalez:2007}
{Gonzalez}, J.~F., {Hubrig}, S., \& {Savanov}, I. 2007, Informational Bulletin
  on Variable Stars, 5794, 1

\bibitem[{{Hatzes} \& {Mkrtichian}(2004)}]{hatzes:2004}
{Hatzes}, A.~P. \& {Mkrtichian}, D.~E. 2004, \mnras, 351, 663

\bibitem[{{Horne} \& {Baliunas}(1986)}]{horne:1986}
{Horne}, J.~H. \& {Baliunas}, S.~L. 1986, \apj, 302, 757

\bibitem[{{Kochukhov}(2008)}]{kochukhov:2008a}
{Kochukhov}, O. 2008, \aap, 483, 557

\bibitem[{{Kochukhov} \& {Bagnulo}(2006)}]{kochukhov:2006}
{Kochukhov}, O. \& {Bagnulo}, S. 2006, \aap, 450, 763

\bibitem[{{Kochukhov} {et~al.}(2002){Kochukhov}, {Landstreet}, {Ryabchikova},
  {Weiss}, \& {Kupka}}]{kochukhov:2002}
{Kochukhov}, O., {Landstreet}, J.~D., {Ryabchikova}, T., {Weiss}, W.~W., \&
  {Kupka}, F. 2002, \mnras, 337, L1

\bibitem[{{Kochukhov} \& {Ryabchikova}(2001)}]{kochukhov:2001b}
{Kochukhov}, O. \& {Ryabchikova}, T. 2001, \aap, 374, 615

\bibitem[{{Kochukhov} {et~al.}(2008){Kochukhov}, {Ryabchikova}, {Bagnulo}, \&
  {Lo Curto}}]{kochukhov:2008b}
{Kochukhov}, O., {Ryabchikova}, T., {Bagnulo}, S., \& {Lo Curto}, G. 2008,
  \aap, 479, L29

\bibitem[{{Kochukhov} {et~al.}(2007){Kochukhov}, {Ryabchikova}, {Weiss},
  {Landstreet}, \& {Lyashko}}]{kochukhov:2007}
{Kochukhov}, O., {Ryabchikova}, T., {Weiss}, W.~W., {Landstreet}, J.~D., \&
  {Lyashko}, D. 2007, \mnras, 376, 651

\bibitem[{{Kuerster} {et~al.}(1997){Kuerster}, {Schmitt}, {Cutispoto}, \&
  {Dennerl}}]{kuerster:1997}
{Kuerster}, M., {Schmitt}, J.~H.~M.~M., {Cutispoto}, G., \& {Dennerl}, K. 1997,
  \aap, 320, 831

\bibitem[{{Kupka} {et~al.}(1999){Kupka}, {Piskunov}, {Ryabchikova}, {Stempels},
  \& {Weiss}}]{kupka:1999}
{Kupka}, F., {Piskunov}, N., {Ryabchikova}, T.~A., {Stempels}, H.~C., \&
  {Weiss}, W.~W. 1999, \aaps, 138, 119

\bibitem[{{Kurtz} {et~al.}(2006){Kurtz}, {Elkin}, {Cunha}, {Mathys}, {Hubrig},
  {Wolff}, \& {Savanov}}]{kurtz:2006b}
{Kurtz}, D.~W., {Elkin}, V.~G., {Cunha}, M.~S., {et~al.} 2006, \mnras, 372, 286

\bibitem[{{Kurtz} {et~al.}(2007){Kurtz}, {Elkin}, \& {Mathys}}]{kurtz:2007a}
{Kurtz}, D.~W., {Elkin}, V.~G., \& {Mathys}, G. 2007, \mnras, 380, 741

\bibitem[{{Maitzen} {et~al.}(2000){Maitzen}, {Paunzen}, {Vogt}, \&
  {Weiss}}]{Maitzen:2000}
{Maitzen}, H.~M., {Paunzen}, E., {Vogt}, N., \& {Weiss}, W.~W. 2000, \aap, 355,
  1003

\bibitem[{{Martinez} \& {Kurtz}(1994)}]{martinez:1994}
{Martinez}, P. \& {Kurtz}, D.~W. 1994, \mnras, 271, 129

\bibitem[{{Mashonkina} {et~al.}(2005){Mashonkina}, {Ryabchikova}, \&
  {Ryabtsev}}]{mashonkina:2005}
{Mashonkina}, L., {Ryabchikova}, T., \& {Ryabtsev}, A. 2005, \aap, 441, 309

\bibitem[{{Mathys} {et~al.}(1997){Mathys}, {Hubrig}, {Landstreet}, {Lanz}, \&
  {Manfroid}}]{mathys:1997b}
{Mathys}, G., {Hubrig}, S., {Landstreet}, J.~D., {Lanz}, T., \& {Manfroid}, J.
  1997, \aaps, 123, 353

\bibitem[{{Mayor} {et~al.}(2003){Mayor}, {Pepe}, {Queloz}, {Bouchy},
  {Rupprecht}, {Lo Curto}, {Avila}, {Benz}, {Bertaux}, {Bonfils}, {dall},
  {Dekker}, {Delabre}, {Eckert}, {Fleury}, {Gilliotte}, {Gojak}, {Guzman},
  {Kohler}, {Lizon}, {Longinotti}, {Lovis}, {Megevand}, {Pasquini}, {Reyes},
  {Sivan}, {Sosnowska}, {Soto}, {Udry}, {van Kesteren}, {Weber}, \&
  {Weilenmann}}]{mayor:2003}
{Mayor}, M., {Pepe}, F., {Queloz}, D., {et~al.} 2003, The Messenger, 114, 20

\bibitem[{{Mkrtichian} {et~al.}(2003){Mkrtichian}, {Hatzes}, \&
  {Kanaan}}]{mkrtichian:2003}
{Mkrtichian}, D.~E., {Hatzes}, A.~P., \& {Kanaan}, A. 2003, \mnras, 345, 781

\bibitem[{{Moon} \& {Dworetsky}(1985)}]{moon:1985}
{Moon}, T.~T. \& {Dworetsky}, M.~M. 1985, \mnras, 217, 305

\bibitem[{{Napiwotzki} {et~al.}(1993){Napiwotzki}, {Schoenberner}, \&
  {Wenske}}]{napiwotzki:1993}
{Napiwotzki}, R., {Schoenberner}, D., \& {Wenske}, V. 1993, \aap, 268, 653

\bibitem[{{Pojmanski}(2002)}]{pojmanski:2002}
{Pojmanski}, G. 2002, Acta Astronomica, 52, 397

\bibitem[{{Ryabchikova} {et~al.}(2008){Ryabchikova}, {Kochukhov}, \&
  {Bagnulo}}]{ryabchikova:2008}
{Ryabchikova}, T., {Kochukhov}, O., \& {Bagnulo}, S. 2008, \aap, 480, 811

\bibitem[{{Ryabchikova} {et~al.}(2004){Ryabchikova}, {Nesvacil}, {Weiss},
  {Kochukhov}, \& {St{\"u}tz}}]{ryabchikova:2004a}
{Ryabchikova}, T., {Nesvacil}, N., {Weiss}, W.~W., {Kochukhov}, O., \&
  {St{\"u}tz}, C. 2004, \aap, 423, 705

\bibitem[{{Ryabchikova} {et~al.}(2002){Ryabchikova}, {Piskunov}, {Kochukhov},
  {Tsymbal}, {Mittermayer}, \& {Weiss}}]{ryabchikova:2002}
{Ryabchikova}, T., {Piskunov}, N., {Kochukhov}, O., {et~al.} 2002, \aap, 384,
  545

\bibitem[{{Ryabchikova} {et~al.}(2006){Ryabchikova}, {Ryabtsev}, {Kochukhov},
  \& {Bagnulo}}]{ryabchikova:2006}
{Ryabchikova}, T., {Ryabtsev}, A., {Kochukhov}, O., \& {Bagnulo}, S. 2006,
  \aap, 456, 329

%\bibitem[{{Ryabchikova} {et~al.}(2007{\natexlab{a}}){Ryabchikova}, {Sachkov},
\bibitem[{{Ryabchikova} {et~al.}(2007){Ryabchikova}, {Sachkov},
  {Kochukhov}, \& {Lyashko}}]{ryabchikova:2007b}
{Ryabchikova}, T., {Sachkov}, M., {Kochukhov}, O., \& {Lyashko}, D.
%  2007{\natexlab{a}}, \aap, 473, 907
  2007, \aap, 473, 907

%\bibitem[{{Ryabchikova} {et~al.}(2007{\natexlab{b}}){Ryabchikova}, {Sachkov},
%  {Weiss}, {Kallinger}, {Kochukhov}, {Bagnulo}, {Ilyin}, {Landstreet}, {Leone},
%  {Lo Curto}, {L{\"u}ftinger}, {Lyashko}, \& {Magazz{\`u}}}]{ryabchikova:2007a}
%{Ryabchikova}, T., {Sachkov}, M., {Weiss}, W.~W., {et~al.} 2007{\natexlab{b}},
%  \aap, 462, 1103

\bibitem[{{Sachkov} {et~al.}(2008){Sachkov}, {Kochukhov}, {Ryabchikova},
  {Huber}, {Leone}, {Bagnulo}, \& {Weiss}}]{sachkov:2008}
{Sachkov}, M., {Kochukhov}, O., {Ryabchikova}, T., {et~al.} 2008, \mnras, 389,
  903

\bibitem[{{Saio}(2005)}]{saio:2005}
{Saio}, H. 2005, \mnras, 360, 1022

\bibitem[{{Vogt} \& {Faundez}(1979)}]{vogt:1979}
{Vogt}, N. \& {Faundez}, A.~M. 1979, \aaps, 36, 477

\end{thebibliography}

\end{document}